\documentclass[aps,prc,reprint,superscriptaddress,nofootinbib]{revtex4-1}

 \usepackage{graphicx}
\usepackage{bm}
\usepackage{amsmath} \usepackage{braket} \usepackage{epsfig} 
\usepackage{tensor}
\usepackage[version=4]{mhchem}
\usepackage{titlesec}
\usepackage{ifthen}
\usepackage[misc]{ifsym}
\usepackage{sidecap}
\usepackage{listings}
\usepackage[para,online,flushleft]{threeparttablex}

\usepackage{hyperref}
\usepackage[all]{hypcap}

\usepackage{float}

\usepackage{multibib}
\newcites{meth}{References}

\renewcommand{\vec}[1]{\mbox{\boldmath $#1$}}

\usepackage{xcolor}

\begin{document}
\title{The Hidden Variables: Harnessing Half-Shell Potentials for Enhanced Precision in Nuclear Reaction Calculations}


\author{Hao Liu}


\affiliation{School of Physics Science and Engineering, Tongji University, Shanghai 200092, China.}

\author{Jin Lei}
\email[Corresponding author: ]{jinl@tongji.edu.cn}

\affiliation{School of Physics Science and Engineering, Tongji University, Shanghai 200092, China.}

\author{Zhongzhou Ren}

\affiliation{School of Physics Science and Engineering, Tongji University, Shanghai 200092, China.}

\begin{abstract}
We explore the impact of half-shell components on nuclear reaction calculations, focusing on nonelastic breakup cross sections within the Ichimura-Austern-Vincent (IAV) model. By advocating for the use of a consistent Single Folding Model (SFM) for all optical potentials in IAV calculations, we aim to reduce the uncertainties associated with half-shell components and enhance agreement with experimental data.
We present results from deuteron-induced reactions on $^{60}$Ni and $^{208}$Pb, which serve as surrogate targets for neutron-induced reactions on short-lived nuclei. The application of consistent optical potentials derived from the SFM shows improved alignment with experimental data compared to traditional global phenomenological potentials. Furthermore, we investigate the $^{59}$Co($^6$Li,$\alpha X$) reaction, which reveals that the half-shell $T$-matrix plays a pivotal role in accurately modeling nuclear reactions. Our findings suggest that a unified approach to optical potentials, accounting for half-shell effects, is critical for a precise understanding of complex nuclear reactions. This work highlights the significance of the internal dynamics of the wave function, particularly in lighter targets, and underscores the importance of the half-shell $T$-matrix as a previously underappreciated variable in reaction calculations.
\end{abstract}


\date{\today}%
\maketitle

\section{Introduction}
The optical potential is a fundamental concept in the study of nuclear reactions, offering a method to describe the complex interactions between a projectile and its target. In practice, the optical potential is often represented by a Woods-Saxon shape with parameters that are adjusted to align with experimental data. This fitting process effectively calibrates the on-shell component of the optical potential, corresponding to situations where the projectile's momentum $k$ remains unchanged in magnitude during the collision. Half-shell components complicate the nucleus-nucleus interaction models. They are crucial in few-body nuclear systems, exemplified by the Phillips line, which connects triton binding energy with neutron-deuteron scattering length, revealing the sensitivity of three-nucleon systems to the details of nucleon-nucleon (NN) interactions~\cite{Faddeev,AGS,hadizadeh2020three,PHILLIPS1968209,PhysRevC.68.034002}. Polyzou and Glöckle demonstrated that half-shell NN interactions, including three-nucleon forces (3NFs), can replicate the effects of distinct NN interactions~\cite{Polyzou1990}.

Understanding and accounting for these half-shell components are therefore essential for accurate theoretical descriptions of nuclear systems, particularly when extending beyond the simplest two-body interactions. In this paper, we investigate the impact of half-shell components on nuclear reaction calculations, focusing particularly on nonelastic breakup (NEB) cross sections. We employ the IAV (Ichimura, Austern, and Vincent) model~\cite{Ichimura1985}, a theoretical framework crafted to unravel the intricacies of NEB, a phenomenon occurring when a composite two-body projectile ($a=b+x$) engages with a target $A$, leading to the detection of one projectile fragment while the other engages nonelastically with the target. This interaction may entail excitations, particle exchanges, or fragment absorption by the target. The IAV model, leveraging the Distorted Wave Born Approximation (DWBA), adeptly addresses the complexities of such NEB events and has shown efficacy in reactions induced by weakly bound projectiles~\cite{Potel2017, Carlson2016,Jin15,Jin15b,Jin17,Jin18,Jin19}.

In the practical application of the IAV model, different optical potentials are required for each pair system involved. The inconsistent use of these optical potentials within a single IAV model calculation can introduce additional systematic errors due to varying behaviors in the interior part of the scattering wave function or differing half-shell properties of the $T$-matrix.

To address these issues, we propose the use of a consistent single folding model as the starting point for all optical potentials in the IAV calculations. This model, which uses the nucleon-A interaction of KD02~\cite{KONING2003231,Lu23}, is applied to deuteron and $^6$Li induced reaction systems. We argue that this approach can reduce the uncertainties associated with half-shell components. Establishing a consistent baseline for the optical potentials minimizes the systematic discrepancies that arise from different interior wave function behaviors and half-shell $T$-matrix properties. Our results demonstrate that this consistent single folding potential approach leads to a better agreement with experimental data, reinforcing the importance of a unified framework in complex nuclear reaction analysis.

The structure of the paper is as follows: Section~\ref{sec:II} introduces the importance of the half-shell component and provides a thorough derivation of the IAV model. In Section~\ref{sec:III}, we apply the formalism to inclusive reactions induced by deuterons and $^6$Li. Lastly, Section~\ref{sec:IV} summarizes the main findings of this study and provides an overview of potential future developments.

\section{Theoretical framework}
\label{sec:II}
\begin{figure*}[tb]
\begin{center} 
 {\centering \resizebox*{1.7\columnwidth}{!}{\includegraphics{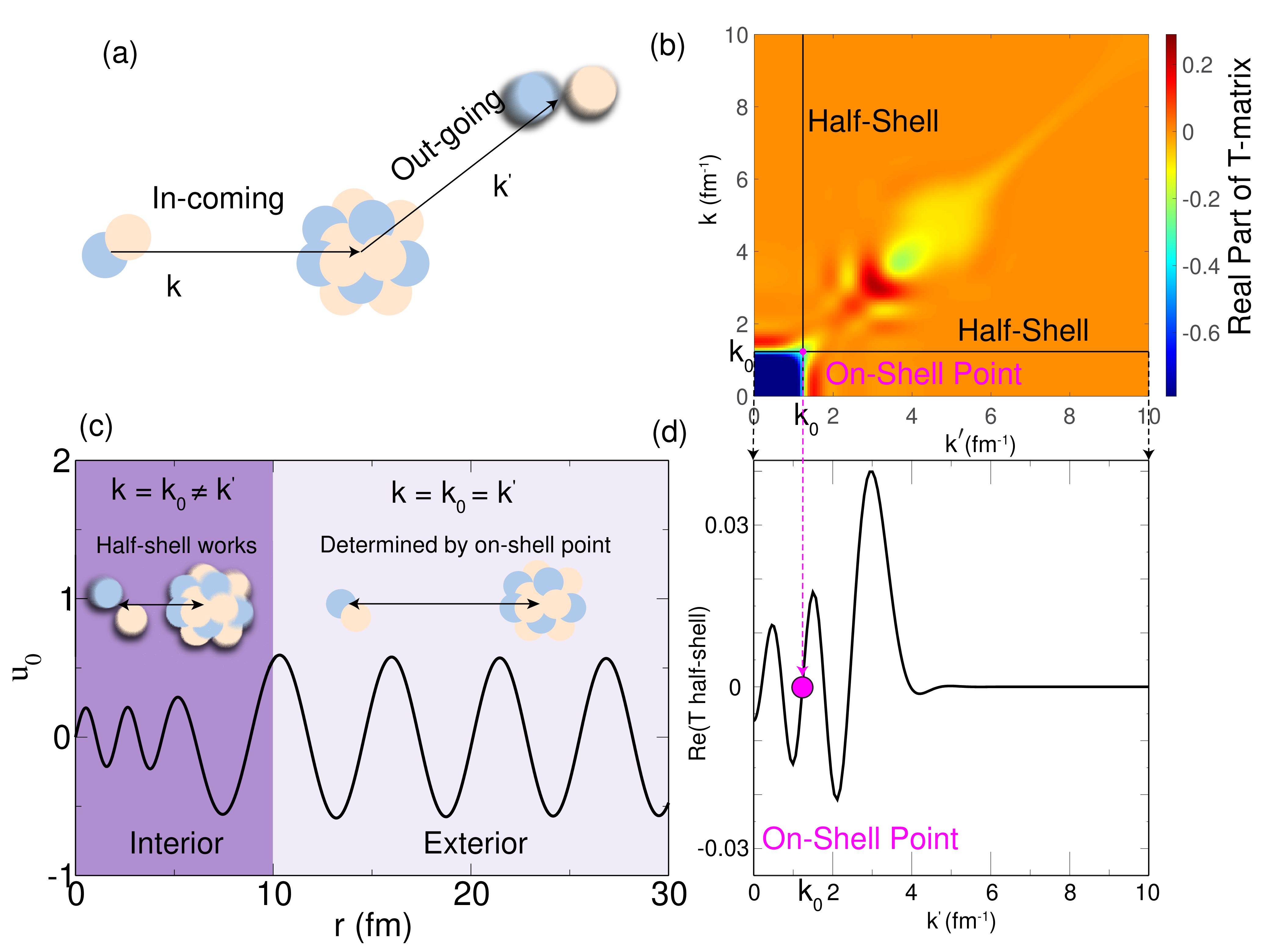}} \par}
\caption{\label{fig:0} Nuclear scattering dynamics and two-body interactions: (a) Elastic scattering of a deuteron with initial momentum $k$ off a target nucleus, resulting in an outgoing momentum $k'$. (b) A contour plot illustrating the real part of the two-body scattering T-matrix, with the half-shell region delineated by the intersection of two solid black lines, and the on-shell point is located at their intersection. This on-shell point dictates the asymptotic behavior of the wave function. (c) The two-body scattering wave function is influenced by the on-shell point at larger inter-nuclear distances and undergoes changes due to half-shell interactions as the nuclei approach one another. (d) A focused representation of the half-shell region from (b), highlighting its impact on the scattering process.}
\end{center}
\end{figure*}

\subsection{Half-shell $T$-matrix and scattering wave function}
In this subsection, we explore the properties of the half-shell $T$-matrix and elucidate its relation to the scattering wave function. The $T$-matrix, a fundamental component in quantum mechanics, satisfies the Lippmann-Schwinger (LS) equation~\cite{QM2}:
\begin{align}
T(\vec{k},\vec{k}',k_0) & = V(\vec{k},\vec{k}',k_0) \nonumber\\
& + \int d^3k'' V(\vec{k},\vec{k}'',k_0) \frac{1}{E^+ - \frac{{{k}''}^2}{2\mu}} T(\vec{k}'',\vec{k}',k_0),  
\end{align}
where $E^+=\lim_{\epsilon\to 0}\frac{k_0^2}{2\mu}+i\epsilon$ and $k_0$ denotes the on-shell point. The term $V(\vec{k},\vec{k}',k_0)$ stands for $\langle \vec{k} | V(k_0)|\vec{k}'\rangle$, indicating an energy-dependent potential. The half-shell $T$-matrix is defined by
\begin{equation}
T(\vec{k},\vec{k}_0,k_0) = \langle \vec{k}| T(k_0) | \vec{k}_0\rangle = \langle \vec{k} | V(k_0)|\chi^{(+)} \rangle. 
\end{equation}
In a pure two-body scenario, only the on-shell $T$-matrix $T(\vec{k}_0,\vec{k}_0,k_0)$ is utilized to calculate observables, as the half-shell part accounts for the internal dynamics of the colliding entities. Consider a scenario where the projectile is a deuteron; the optical potential must encapsulate the potential for the deuteron to be excited into its continuum post-collision, which may result in a change in the magnitude of the outgoing momentum compared to the incoming one, as depicted in Fig.~\ref{fig:0} (a). The shadow of the deuteron indicates that after the collision, the deuteron may get excited into its continuum, and therefore the magnitude of outgoing momentum differs from the incoming one. These half-shell components act as hidden variables that are not directly observable in two-body elastic scattering but become pivotal in reactions involving greater degrees of freedom.

However, the half-shell $T$-matrix $T(\vec{k},\vec{k}_0,k_0)$, depicted in Fig.~\ref{fig:0} (d) and representing the half-shell part of the whole $T$-matrix as shown in Fig.~\ref{fig:0} (b), is for the $s$-wave reaction system of $d+^{60}$Ni at an incident energy of 17 MeV in the Lab frame. The potential used in the calculations is taken from Ref.~\cite{YYQ06}. This half-shell $T$-matrix can be used to compute the scattering wave function, given by
\begin{equation}
\langle \vec{r} | \chi^{(+)} \rangle = \langle \vec{r} | \vec{k}_0 \rangle + \int d\vec{k} \langle \vec{r} | \vec{k} \rangle G_0^{(+)}(k) T(\vec{k},\vec{k}_0,k_0),
\end{equation}
where $\langle \vec{r} | \vec{k}\rangle$ is the plane wave. This relationship indicates that the half-shell $T$-matrix is integral to deriving the scattering wave function, with the on-shell point determining the asymptotic behavior and the off-shell points affecting the interior of the wave function.

As depicted in Fig.~\ref{fig:0} (c), the characteristics of the scattering wave function can be explored by computing the half-shell $T$-matrix. On-shell points, defined by $k = k' = k_0$, dictate the asymptotic behavior of the scattering wave function, as represented by the $S$-matrix. The half-shell components, denoted by $k = k_0 \neq k'$, play a crucial role in the detailed structure of the wave function within the interaction region. This highlights the significance of half-shell interactions in shaping the wave function. Consequently, they potentially impact nuclear reaction outcomes, especially in scenarios that require consideration of more complex internal structures and degrees of freedom.

\subsection{The IAV model}
We briefly review the IAV model~\cite{AUSTERN1987125,Ichimura1985} here.
The inclusive breakup reaction under study is described by the equation
\begin{equation}
    a(=b+x)+A \rightarrow b+ B^{*},
    \label{eq1}
\end{equation}
where the projectile $a$ has a two-body structure $(b+x)$, $b$ is the detected particle, and $B^{*}$ denotes any possible final state of the $x+A$ system. In the IAV model, the fragment $b$ is referred to as the spectator, while the fragment $x$ is considered the participant.

The IAV model provides the NEB cross section as
\begin{equation}
   \frac{\mathrm{d}^2\sigma}{\mathrm{d}\Omega_b \mathrm{d}E_b}\Big|_{\textbf{post}}^{\textbf{NEB}} = -\frac{2}{\hbar v_a}\rho_b(E_b) \langle  \psi_x (\boldsymbol{k}_b) | W_x  |\psi_x (\boldsymbol{k}_b) \rangle,
    \label{eq:IAV}
\end{equation}
where $v_a$ is the projectile-target relative velocity, $\rho_b(E_b)=\frac{\mu_b k_b}{(2\pi)^3 \hbar^2}$ is the density of states for particle $b$, with $\mu_b$ and $k_b$ being the reduced mass and wave number, respectively. The term $W_x$ represents the imaginary part of $U_{x}$, which characterizes the $x+A$ elastic scattering, and $\psi_x$ is the $x$-channel wave function obtained by solving the inhomogeneous differential equation
\begin{equation}
    (E_x-K_x-U_{x})\psi_x(\boldsymbol{k}_b,\boldsymbol{r}_x)
    =
    \langle \boldsymbol{r}_x
    \chi_b^{(-)}(\boldsymbol{k}_b)
    |V_{\mathrm{post}}|
    \chi_a^{(+)}\phi_a\rangle,
    \label{eq:inhomo}
\end{equation}
where $E_x = E - E_b$. $K_x$ is the kinetic energy operator for the relative motion between fragment $x$ and target $A$, $\chi_b^{(-)}$ is the scattering wave function with incoming boundary conditions, describing the scattering of $b$ in the final channel with respect to the $x+A$ subsystem. The post-form transition operator $V_{\mathrm{post}} = V_{bx} + U_{bA} - U_{bB}$ includes $V_{bx}$, the potential binding the clusters $b$ and $x$ in the initial composite nucleus $a$, $U_{bA}$, the fragment-target optical potential, and $U_{bB}$, the optical potential in the final channel. Furthermore, $\chi_a^{(+)}$ is the distorted wave describing the $a+A$ elastic scattering with an outgoing boundary condition, and $\phi_a$ is the initial ground state of the projectile $a$.

\section{Result}
\label{sec:III}
\subsection{Application to $(d,pX)$}
In the present research, we focus on deuteron-induced reactions $(d,pX)$ under standard kinematic conditions, employing isotopes $\mathrm{^{60}Ni}$ and $\mathrm{^{208}Pb}$ as target nuclei. To facilitate reliable comparisons between different systems and to reduce systematic errors, we have chosen incident energies that align with the peaks of the Coulomb barriers for each nucleus. Specifically, we analyzed the $\mathrm{^{60}Ni}(d,pX)$ reaction at an incident energy of 23 MeV and the $\mathrm{^{208}Pb}(d,pX)$ reaction at 55 MeV, with both energies referenced in the laboratory frame. The experimental results pertaining to the $\mathrm{^{60}Ni}(d,pX)$ reaction have been previously reported and are available for review in the referenced literature~\cite{expdpx}.

To underscore the impact of the effective $d+A$ interaction in IAV model, we meticulously evaluate and compare the results derived from employing the Single Folding Model (SFM) with the KD02 global nucleon-target interaction against those obtained using two sets of established global phenomenological optical model potentials for deuteron-induced reactions: the Han-Shi-Shen (HSS)~\cite{YYQ06} and the An-Cai (AC)~\cite{AnCai} potentials. Due to a lack of experimental data, the renormalization factors of SFM potential were determined by fitting to the results of elastic scattering data obtained through the Continuum Discretized Coupled Channels (CDCC) method and experimental data~\cite{TAKEI198741, Matsuoka:1986hpk}. For the $d-\mathrm{Ni}$ system, we determined the real part to be $Nr=0.669$ and the imaginary part as $Ni=1.125$. Similarly, for the $d-\mathrm{Pb}$ system, the real and imaginary parts were found to be $Nr=0.899$ and $Ni=1.169$, respectively.

\begin{figure}[bt]
\begin{center}
 {\centering \resizebox*{1.0\columnwidth}{!}{\includegraphics{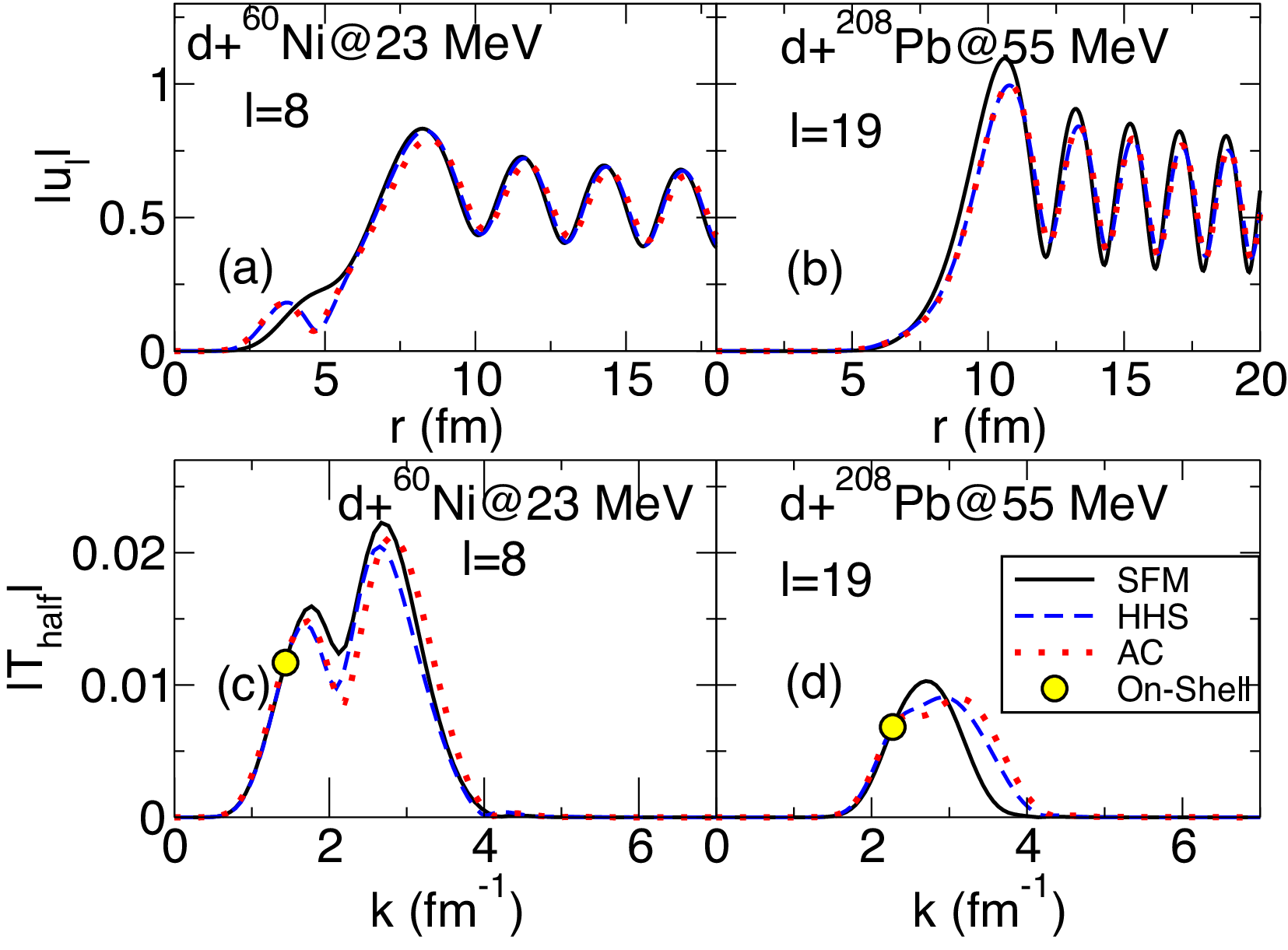}} \par}
\caption{\label{fig:2} The absolute value of the radial wave functions and the half-shell $T$-matrix ($T_{\text{half-shell}}$) for $d+\mathrm{^{60}Ni}$ at an incident energy of 23 MeV in the Lab frame for $l=8$ are depicted in Panel (a) and Panel (c), respectively. Similarly, the absolute value of the radial wave functions and the half-shell $T$-matrix for $d+\mathrm{^{208}Pb}$ at an incident energy of 55 MeV in the Lab frame for $l=19$ are plotted in Panel (b) and Panel (d), respectively. Calculated results with SFM, HSS, and AC potentials are depicted as solid black lines, dashed blue lines, and dotted red lines, respectively, with on-shell points indicated by yellow circles.}
\end{center}
\end{figure}

To elucidate the influence of these diverse optical potentials within the IAV model, we showcase the absolute value of radial components of the scattering wave function for $l=8$, corresponding to the relative angular momentum between $d+\mathrm{^{60}Ni}$, in Fig.~\ref{fig:2} (a), and for $l=19$, correlating with the relative angular momentum between $d+\mathrm{^{208}Pb}$, in Fig.~\ref{fig:2} (b). These particular values of angular momentum were selected due to their predominant contribution to the NEB cross section. The findings and a more detailed discussion on this matter are presented subsequently. The scattering wave functions are depicted using solid lines for the SFM results, dashed lines for the HSS potential, and dotted lines for the AC potential, allowing for a clear comparison of the implications these potentials have.

Observations from Fig.~\ref{fig:2}(a) and (b) confirm that all the optical potentials under consideration are phase-equivalent, indicating their equal aptitude in describing elastic scattering data. However, notable differences are discernible in the inner regions of the wave functions. To delve deeper into these variations, we plot the absolute value of the half-shell $T$-matrix for the same reaction system and partial waves in Fig.~\ref{fig:2} (c) for $d+\mathrm{^{60}Ni}$ and Fig.~\ref{fig:2} (d) for $d+\mathrm{^{208}Pb}$. The half-shell $T$-matrix is a critical component in the IAV calculation, accounting for off-energy-shell effects that are pivotal for the comprehension of breakup reactions where the deuteron disintegrates, leading to the relative energy of the $np$ pair with respect to the target being off the energy shell. Furthermore, the half-shell $T$-matrix is instrumental in calculating the scattering wave function through the Lippmann-Schwinger equation. It is evident that the on-shell $T$-matrix calculated with these different potentials concurs, while the left part exhibits a marked divergence. The on-shell point is related to the scattering matrix, which is used to generate the scattering wave function asymptotically, whereas the left part of the half-shell $T$-matrix contributes to the internal part of the wave function, which is the region where the deuteron interacts nonelastically with the target.

\begin{figure}[tb]
\begin{center}
 {\centering \resizebox*{1.0\columnwidth}{!}{\includegraphics{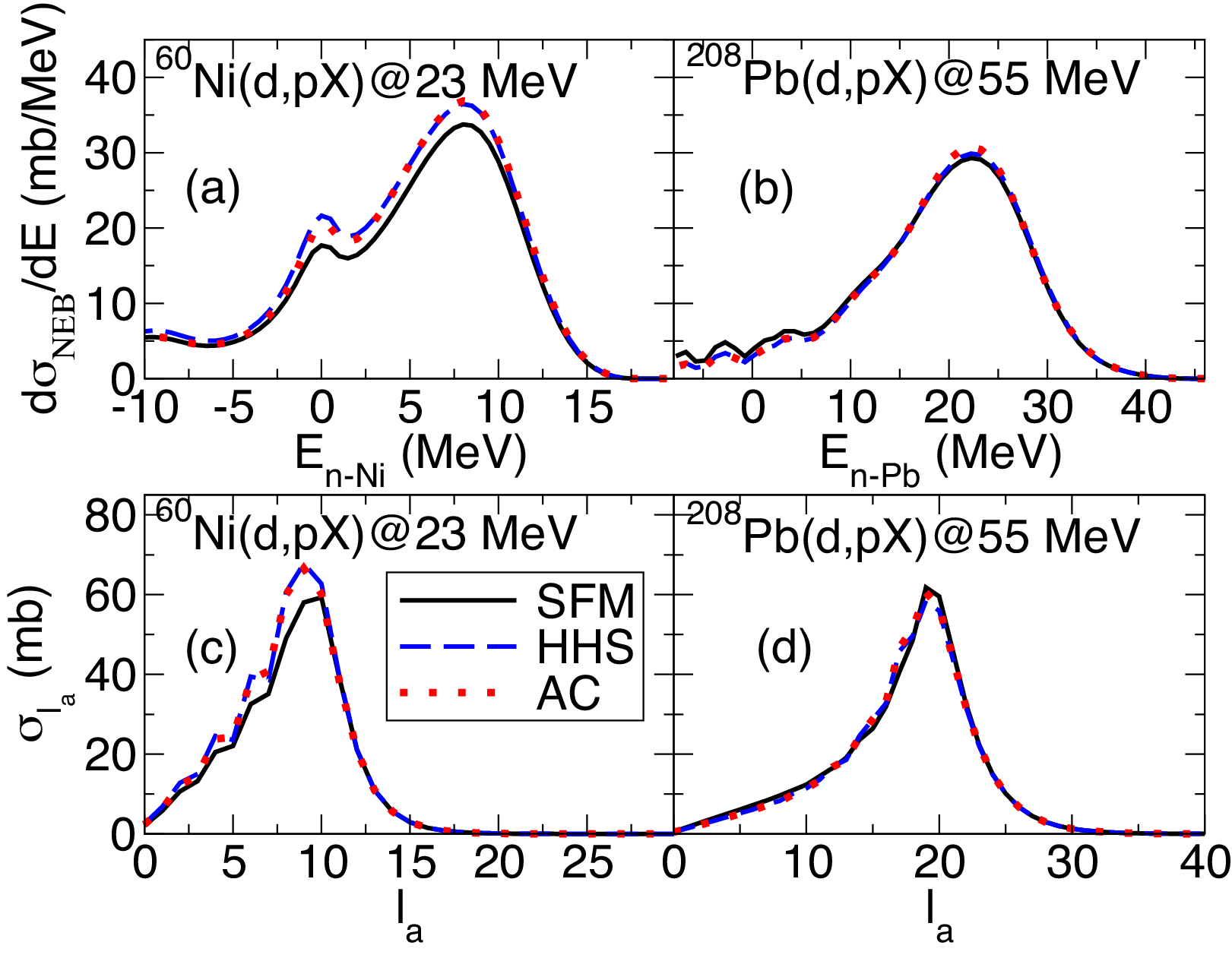}} \par}
\caption{\label{fig:1}Panels (a) and (b) display the differential cross-section of NEB energy distribution with respect to the relative energies between $n-{}^{60}\text{Ni}$ and $n-{}^{208}\text{Pb}$ targets for the reaction systems $^{60}$Ni($d$,$pX$) at 23 MeV and $^{208}$Pb($d$,$pX$) at 55 MeV, respectively. Panels (c) and (d) explore the contributions of the partial wave $l_a$, which is the relative partial wave between the projectile and target, to the NEB cross sections for the reaction systems $^{60}$Ni($d$,$pX$) at 23 MeV and $^{208}$Pb($d$,$pX$) at 55 MeV, respectively. The black solid lines represent the results obtained with the SF potential, the blue dashed lines represent the results obtained with the HSS potential, and the red dotted lines represent the results obtained with the AC potential.}
\end{center}
\end{figure}

To explore these half-shell effects within the IAV model, we computed the NEB cross section of the $(d,pX)$ reaction for targets $\mathrm{^{60}Ni}$ and $\mathrm{^{208}Pb}$. The differential cross-section energy distributions with respect to the relative energies between $n-\mathrm{^{60}Ni}$ and $n-\mathrm{^{208}Pb}$ are depicted in Fig.~\ref{fig:1} (a) and (b), respectively. The solid, dashed, and dotted lines correspond to the results obtained with the SFM, HSS, and AC optical potentials for $d+$target, respectively. The $n-p$ relative potential was modeled using a simple Gaussian form~\cite{AUSTERN1987125}, and the left interactions in the IAV model were sourced from KD02~\cite{KONING2003231}. Notably, in both reactions, the HSS and AC potentials closely align with each other but deviate from the SFM potential results for the $\mathrm{^{60}Ni}(d,pX)$ reaction. Conversely, all three potentials show good agreement for the $\mathrm{^{208}Pb}(d,pX)$ reaction. These observations are consistent with the hypothesis that for heavier targets, the surface approximation is sufficient and only the asymptotic part affects the NEB cross sections~\cite{Junzhe23}. However, for lighter targets, the internal part of the scattering wave function plays a more critical role.

In an effort to further understand this distinction, we display the integrated NEB cross section as a function of angular momentum in Fig.~\ref{fig:1} (c) for $\mathrm{^{60}Ni}(d,pX)$ and Fig.~\ref{fig:1} (d) for $\mathrm{^{208}Pb}(d,pX)$. It is evident that for the heavier target, all the results concur across the entire range of partial waves, whereas for the lighter target, discrepancies emerge in the lower partial waves. This indicates that the surface approximation holds validity for higher partial waves, but for lower partial waves, particularly in lighter targets, the internal dynamics of the wave function are more influential.

\begin{figure}[tb]
\begin{center}
 {\centering \resizebox*{0.85\columnwidth}{!}{\includegraphics{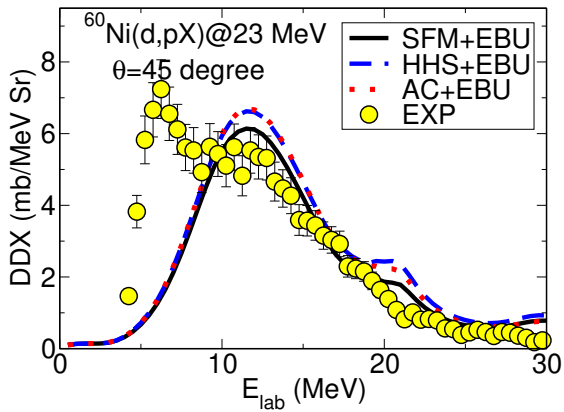}} \par}
\caption{\label{fig:7}The double-differential cross section (DDX), \(d^2\sigma / (dE_p d\Omega_p)\), as a function of outgoing proton energy in the \( \mathrm{^{60}Ni}(d,pX) \) reaction at 23 MeV and at an outgoing proton angle of \( 45^\circ \) in the Lab frame, is presented. The yellow points represent the experimental data~\cite{expdpx}. The solid black line is the sum of the NEB and EBU cross sections calculated using the SFM; the blue dashed line shows the sum calculated using the HSS potential~\cite{YYQ06}; and the red dotted line represents the sum calculated using the AC potential~\cite{AnCai}.}
\end{center}
\end{figure}

Lastly, we compare the calculated inclusive breakup cross section with experimental data. Fig.~\ref{fig:7} displays the double differential cross-section $d^2\sigma / (dE_pd\Omega_{p})$ as a function of the outgoing proton energy in the lab frame for the reaction $\mathrm{^{60}Ni}(d,pX)$ at 23 MeV, with the proton detected at $\theta=45^\circ$. The elastic breakup (EBU) is computed by the Continuum Discretized Coupled Channels (CDCC) method, treating the deuteron breakup as inelastic excitations to the $p-n$ continuum, which is discretized in energy bins. For this case, $p-n$ states up to $\ell=0-2$ partial waves and a maximum excitation energy of 18 MeV were included. The combined EBU and NEB results with different $d+\mathrm{^{60}Ni}$ optical potentials show a bell-shaped distribution, peaking around half the deuteron energy. However, the sum of these contributions does not account for the experimental yield at low proton energies, which primarily result from compound nucleus processes followed by evaporation and pre-equilibrium, not covered by our formalism. Results obtained with the SFM potential demonstrate improved agreement with the experimental data~\cite{expdpx} compared to those obtained with the HHS and AC potentials. It is noteworthy that using the IAV model with the CDCC wave function can inherently include the correct half-shell properties in the $ d-\text{target} $ channel, as discussed in Ref.~\cite{Jin23}.

\subsection{Application to $(\mathrm{^6Li},\alpha X)$.}
\begin{figure}[tb]
\begin{center}
 {\centering \resizebox*{0.85\columnwidth}{!}{\includegraphics{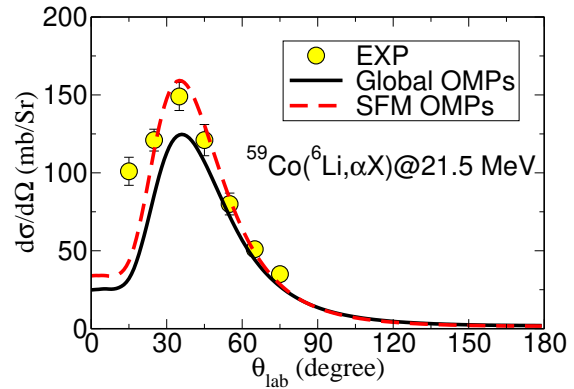}} \par}
\caption{\label{fig:8} The differential cross section angular distribution for the \( \mathrm{^{59}Co(^{6}Li}, \alpha X) \) reaction at 21.5 MeV is depicted. The yellow points correspond to the experimental data~\cite{SOUZA200936}. The solid black line illustrates the results obtained using the global phenomenological optical model potential, and the red dashed line indicates the results using the full SFM potential.
}
\end{center}
\end{figure}
As a second example, we turn our attention to the production of alpha particles following the breakup of the weakly bound nucleus $^6$Li. We examine the reaction $^{59}$Co($^6$Li,$\alpha X$) at an incident energy of 21.5 MeV in the laboratory frame. The experimental data is taken from Ref.~\cite{SOUZA200936}. To streamline our calculations, we have chosen to disregard the internal spin of the involved particles. For the interaction between the $\alpha$ particles and the $d$, we employ a Woods-Saxon potential characterized by parameters: depth $V_0 = 78.46$ MeV, radius $r_0 = 1.15$ fm, and diffuseness $a = 0.7$ fm, as outlined in Ref.~\cite{NISHIOKA1984230}.

In our analysis, we consider two distinct sets of optical potentials. The first set utilizes global optical potential parameters for the interactions between $^6$Li and $^{59}$Co, between $d$ and $^{59}$Co, and between $\alpha$ and $^{59}$Co, as reported in Refs.~\cite{COOK1982153,YYQ06,sh15}, respectively. For the second set, we apply the SFM optical potentials to all interactions involved in both the CDCC and the IAV calculations. To ensure that the SFM optical potentials reproduce the same phase shifts as the global optical potentials, normalization factors for the SFM potentials are employed. The normalization factor of $\mathrm{^{6}Li}$ is taken from Ref.~\cite{Lu23}. For the deuteron, we use $Nr=0.899$ and $Ni=0.905$. For the $\alpha$ potential, we use $Nr=1.34$ and $Ni=2.0$.

The differential cross section as a function of the outgoing $\alpha$ particle angles in the laboratory frame is depicted in Fig.~\ref{fig:8}. The solid line represents the combined results of the EBU and NEB using the global optical potential parameter sets, while the dashed line corresponds to the combined results of EBU and NEB with fully SFM optical potentials. It is evident that the results utilizing SFM optical potentials significantly improve the concordance with the experimental data. These findings suggest that the half-shell $T$-matrix, which cannot be directly constrained by two-body scattering data, serves as a hidden variable that can enhance the precision of reaction calculations. This underlines the importance of the half-shell $T$-matrix in capturing the complexities of the reaction dynamics, thereby providing a more accurate depiction of the processes involved.

\section{Discussion}
\label{sec:IV}
Our investigations into the role of half-shell components in nuclear reaction calculations have provided significant insights into the NEB cross sections within the IAV model. In particular, the NEB mechanism is crucial for investigating reaction processes such as knockout and surrogate reactions. Due to numerical limitations, most of these processes are based on the DWBA. By employing a consistent SFM for all optical potentials in the IAV calculations, we have demonstrated a marked improvement in alignment with experimental data. This suggests that the half-shell components of the interaction and the interior behavior of the resultant wave functions are critical to accurately characterizing NEB processes.

The traditional approach of fitting optical potentials to experimental data does not fully capture the complexity of nuclear reactions. Our results highlight the limitations of this approach, especially when discrepancies in half-shell properties can lead to significant systematic errors in NEB cross sections. The consistent use of SFM optical potentials, derived from the KD02 nucleon-target interaction, mitigates these errors, providing a more robust framework for analyzing complex reactions.

Our study also highlighted the importance of lower partial waves in the integrated NEB cross section for lighter targets. The discrepancies observed in these waves between different optical potentials reinforce the need for a more detailed understanding of the internal wave function dynamics to accurately predict reaction outcomes.

In conclusion, our work presents a compelling case for the adoption of consistent SFM optical potentials in NEB cross section calculations within the IAV model. The enhanced agreement with experimental data achieved through this approach underlines the necessity of a unified and comprehensive modeling of optical potentials. Future research should focus on further validating this approach across a broader range of targets and projectiles, as well as integrating these findings into the development of more sophisticated and predictive nuclear reaction models.

\begin{acknowledgments}
We are grateful to Pierre Descouvemont for a critical reading of the manuscript.
This work has been supported by National Natural Science Foundation of China (Grants No.12105204 and No.12035011), by the National Key R\&D Program of China (Contracts No. 2023YFA1606503), and by the Fundamental Research Funds for the Central Universities.
\end{acknowledgments}
\bibliography{reference.bib}

\end{document}